\documentclass[twocolumn,showpacs,preprintnumbers,amsmath,amssymb,prb]{revtex4}

\usepackage{graphicx}% Include figure files
\usepackage{epsfig}% Include figure files
\usepackage{dcolumn}% Align table columns on decimal point
\usepackage{bm}% bold math

\def\sp#1{$^{#1}$}
\def\etal{{\it et al.}}
\def\ccto{CaCu$_3$Ti$_4$O$_{12}$}
\def\im3{{Im\overline 3}}

\begin{document}

\title{Temperature dependent total scattering structural study of CaCu$_3$Ti$_4$O$_{12}$}

\author{E. S. Bo\v zin,$^1$ V. Petkov,$^2$ P. W. Barnes,$^3$ P. M. Woodward,$^3$
T. Vogt,$^4$ S. D. Mahanti,$^1$ and S. J. L. Billinge$^1$}
\affiliation{$^1$Department of Physics and Astronomy and Center for
Fundamental Materials Research, Michigan State University, Biomedical Physical
Sciences, East Lansing, MI 48824-2320.\\
$^2$Physics Department, Central Michigan University, Mount Pleasant,
MI 48859.\\
$^3$Department of Chemistry, Ohio State University, Columbus, OH 43210-1185.\\
$^4$ Physics Department, Brookhaven National Laboratory, Upton, NY 11973-5000.
}

\date{\today}

%\maketitle

\begin{abstract}
X-ray and neutron powder diffraction data as a function of temperature
are analyzed for the colossal dielectric constant material \ccto .
The local structure is studied using atomic pair distribution function
analysis. No evidence is found for enhanced oxygen displacement parameters 
suggesting that short-range octahedral tilt disorder is minimal. However, 
an unusual temperature dependence for the atomic displacement parameters 
of calcium and copper is observed. Temperature dependent modeling of the 
structure, using bond valence concepts, suggests that the calcium atoms 
become underbonded below approximately 260~K, which provides a rationale 
for the unusually high Ca displacement parameters at low temperature.
\end{abstract}
\pacs{61.10.-i,61.12-q,77.22.-d,77.84.-s}
%}
\maketitle
\section{INTRODUCTION}

% set intro

\begin{figure}[tb]
\begin{center}$\,$
\epsfxsize=3.2in
\epsfbox{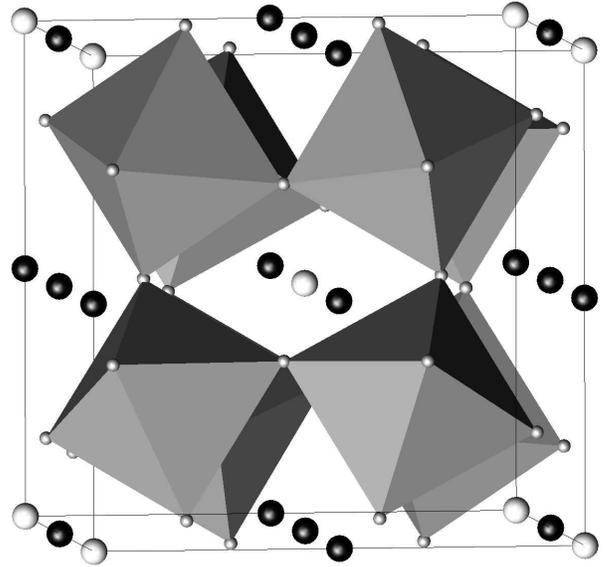}
\end{center}
\caption{CaCu$_3$Ti$_4$O$_{12}$ structure is cubic (space group $\im3$): 
shown are TiO$_6$ octahedra, Ca atoms (large light spheres), and Cu atoms 
(dark spheres).}
\protect\label{fig;structure}
\end{figure}

Recently there has been considerable interest in the dielectric
properties of the cubic perovskite-related CaCu$_3$Ti$_4$O$_{12}$
(CCTO). This material exhibits a giant dielectric constant response
with a highly unusual temperature dependence. It has a high and
relatively temperature independent low-frequency dielectric constant
over a wide temperature range between 100~K and 600~K. However, below 100~K
the value drops abruptly by almost three orders of magnitude,
an effect that is  not accompanied by a long-range structural
phase transition.\cite{subra;jssc00,ramir;ssc00,homes;sci01} The crystal
structure of CCTO (s.g. $\im3$)~\cite{bochu;jssc79} is shown in
Fig.~\ref{fig;structure}.
%%%%%%%%%%%%%%%%%%%%%%%%%%%%%%%%%%%%%%%%%%%%%%%%%%%%%%%%%%%%%%%%%%
%  FIGURE
%%%%%%%%%%%%%%%%%%%%%%%%%%%%%%%%%%%%%%%%%%%%%%%%%%%%%%%%%%%%%%%%%%
%%%%%%%%%%%%%%%%%%%%%%%%%%%%%%%%%%%%%%%%%%%%%%%%%%%%%%%%%%%%%%%%%%
%  FIGURE
%%%%%%%%%%%%%%%%%%%%%%%%%%%%%%%%%%%%%%%%%%%%%%%%%%%%%%%%%%%%%%%%%%
Its structure can be derived from the ideal cubic perovskite structure by
superimposing a body centered ordering of Ca and Cu ions and a pronounced
tilting of the titanium centered octahedra (tilt system
a$^+$a$^+$a$^+$).\cite{glazer;actab72} The tilting dramatically alters
the coordination environments of the A-site cations, leading to a
4-coordinate square-planar environment for Cu and a 12-coordinate
icosahedral environment for Ca.  It is the mismatch in size and bonding
preferences of these two ions and the titanium that drives the large
octahedral tilting distortion. The system is an antiferromagnetic insulator
below T$_N=25$~K, as indicated by transport, neutron diffraction, and Raman
studies.\cite{subra;jssc00,ramir;ssc00,koitz;prb02}

% describe current status

Numerous proposals have been put forward to uncover the
origin of the giant dielectric response in CCTO, including both
intrinsic and extrinsic mechanisms. Intrinsic mechanisms comprise
those referring to a lattice instability resulting in
ferroelectric or relaxor-like behavior, coupled with a highly correlated
electronic ground state. Contrary to other stoichiometric titanates,
no ferroelectric transition has been observed in CCTO. Relaxor-like
dynamical slowing down of dipolar fluctuations in nanosize domains
has been proposed by Homes {\it et al.} from the optical conductivity
measurements.~\cite{homes;sci01} It has been argued that the high dielectric
constant may arise from local dipole moments associated with off-center
displacements of Ti atoms.~\cite{homes;sci01} The structure does not
exhibit a long range structural distortion, as typically observed
in perovskite compounds.  Within this model the absence of a structural
phase transition is explained by the premise that the tilts of the
TiO$_6$ octahedra are large enough to uncouple the local titanium
distortions from each other, effectively decoupling the ferroelectric
order parameter and the long-range crystal structure. The implications
of this model could be tested experimentally using local structural probe, 
such as the atomic pair distribution function analysis~(PDF). The off-center
displacements of Ti would imply large atomic displacement (thermal)
parameters, and possibly octahedral tilt disorder.

Recently, Ramirez and collaborators suggested that local structural defects
within the Ca-Cu sublattice of CCTO, if evidenced, may provide an intrinsic
scenario for explaining the gigantic dielectric constant observed in this
system.\cite{ramir;cm02} In this view, the defects would disrupt an otherwise
rigid Cu-O complex, and would be prone to {\it local} polarizability. However, 
since the defects are random, the polarization does not propagate over long 
range. According to the authors, significant diffuse scattering 
signal is expected in the high-dielectric-constant regime, with freezing at 
lower temperatures. Experimental verification of these predictions could also 
be performed by a local structural study such as EXAFS, or the PDF method that 
utilizes both Bragg and diffuse intensities.

Another possible explanation for the colossal dielectric properties of CCTO
can be an extrinsic mechanism related to the microstructure of the sample,
its morphology, and boundary layer effects.\cite{he;prb02} Subramanian {\it
et al.} argue that unusually high dielectric constant of CCTO is due to its
microstructure because of the creation of an effective circuit of parallel
capacitors as found in boundary-layer dielectrics.\cite{subra;jssc00}
Recent X-ray diffraction study on CCTO single crystals showed twinning on 
a fine scale in the microstructure of CCTO.\cite{subra;sss02} Twin boundaries 
are proposed to act in a manner to create a barrier layer capacitance, the 
effect known to produce extremely high dielectric constants. Sinclair 
{\it et al.} attribute the giant dielectric phenomenon to a grain boundary 
barrier layer capacitance.\cite{sincl;apl02} The model features 
semiconducting perovskite grains with thin, insulating grain boundaries. 
The colossal dielectric constant has also been discussed to possibly 
originate from Maxwell-Wagner type contributions of depletion layers at 
the interface between sample and contacts or at grain boundaries, as has 
quite impressively been demonstrated on a variety of different 
materials.~\cite{lunke;prb02}

%
% describe our contribution

In order to assess the validity of various explanations for the giant
dielectric response of CCTO it is necessary to characterize not only the
average crystal structure, but also the local structure. In this work local
structural properties of the CCTO system are investigated as a function of
temperature within a range from 50~K up to room temperature using atomic pair
distribution function analysis of the powder diffraction 
data.~\cite{warre;bk90,egami;mt90,billi;b;lsfd98}
\section{EXPERIMENTAL}

The CaCu$_3$Ti$_4$O$_{12}$ sample was prepared by conventional
solid state methods. Stoichiometric amounts of CaCO$_3$
(Mallinckrodt, 99.95\% purity), CuO (J.T. Baker, 99.9\%
purity), and TiO$_2$ (EM Science, 99.7\% purity) were weighed and
mixed intimately using acetone in an agate mortar and pestle.
The reaction mixture was preheated in a high form alumina
crucible at 900$^{\circ}$C for 8 hours. The preheated powder
was reground and pressed into pellets. The pellets were
heated in the presence of air at 1075$^{\circ}$C for 24 hours.
Purity of the final product was confirmed by x-ray powder
diffraction using a Bruker D8 Advance diffractometer
equipped with an incident beam Ge monochromator and a
Braun position sensitive detector.

We performed a complementary neutron diffraction and x-ray diffraction study
on powder samples of CaCu$_3$Ti$_4$O$_{12}$. Time-of-flight neutron
diffraction experiments were carried out at Special Environment Powder
Diffractometer (SEPD) at the Intense Pulsed Neutron Source (IPNS) at Argonne
National Laboratory. About 10~g of polycrystalline sample was sealed in a
cylindrical vanadium tube with He exchange gas. The sample was cooled using
a closed-cycle He refrigerator, and data collected at 50~K, 100~K, 150~K,
200~K and 290~K. X-ray diffraction data were collected at X7A beamline of the
National Synchrotron Light Source (NSLS) at Brookhaven National Laboratory.
The sample was carefully packed between Kapton foils to avoid texture
formation, and mounted within a closed cycle helium refrigerator. 
Data were collected at 50~K and 300~K in symmetric transmission geometry 
using 29.08~keV synchrotron radiation ($\lambda = 0.4257$~\AA ). Scattered 
radiation was collected with an intrinsic germanium detector connected to 
a multichannel analyzer. Several runs were conducted and the resulting X-ray 
diffraction patterns were averaged to improve the statistical accuracy and 
reduce any systematic effect due to instabilities in the experimental setup. 
The raw data were background-subtracted, corrected for flux and experimental 
effects such as sample absorption and multiple scattering, and
normalized.\cite{billi;b;lsfd98}

From the normalized data the total scattering structure function,
$S(Q)$,\cite{warre;bk90} is obtained, where
$Q=\arrowvert {\bf Q}\arrowvert =\arrowvert {\bf k}-{\bf k_0}\arrowvert $
represents momentum transfer (magnitude of the scattering vector).
The PDF function, $G(r)$, is then obtained by a Fourier
transformation according to
\begin{equation}
G(r)= \frac{2}{\pi }\int_{0}^{\infty} Q [S(Q)-1] \sin (Qr)\> dQ.
\end{equation}
The strength of the PDF technique is that it takes into account both Bragg
and diffuse scattering components, and yields structural information on both
long-range order and local structural disorder in materials. The function
$G(r)$ gives the number of atoms in a spherical shell of unit thickness at a
distance $r$ from a reference atom. It peaks at characteristic distances
separating pairs of atoms and thus reflects the atomic
structure.\cite{billi;b;lsfd98} Neutron diffraction data processing was
performed using PDFgetN analysis program package.\cite{peter;jac00} The
neutron PDFs examined in this paper use data over a wide momentum
transfer range, and were terminated at $Q_{max}=28$~\AA\sp{-1}.
X-ray diffraction data were processed with help of the programs
RAD~\cite{petko;jac89} and IFO.\cite{petko;jac98} The X-ray diffraction
PDFs investigated here use data up to $Q_{max}=24$~\AA\sp{-1}.

Data have been analyzed using the Rietveld method performed by GSAS 
program.\cite{larso;laur94} Structural information is obtained from 
the PDF data through a modeling procedure similar to the Rietveld
method. However, it is carried out in {\it real-space} and yields the 
{\it local} structure rather than the average crystal 
structure.\cite{billi;b;lsfd98} From the refinements, different 
structural information can be extracted, such as lattice parameters, 
average atomic positions and amplitudes of their thermal motion, atomic 
displacements, and magnitudes of local octahedral tilts. The results 
reported here are obtained using the PDFFIT modeling 
program.\cite{proff;jac99} In cases where the average and local structure
are the same (i.e., well-ordered crystals) the structural parameters obtained
from Rietveld and the PDF are the same.\cite{proff;prb99} However, when
the local structure deviates from the average structure the parameters from
the PDF will reflect the local structure and the Rietveld parameters the
periodically averaged values.

\section{RESULTS}

First we compare our results with published structural data. Several studies
of the average crystalline structure
exist.\cite{subra;jssc00,ramir;ssc00,bochu;jssc79,subra;sss02} None of these
studies finds evidence of a structural distortion as a function of
temperature or pressure that would couple to the dielectric
properties.\cite{ramir;cm02} Despite the possible importance of the lattice, 
these structural studies are not thorough; either being derived solely from
x-ray data (and therefore rather insensitive to important oxygen structural
parameters) or presenting a sparse set of data points.  In addition,
a systematic temperature dependent local structural study has not been
performed, to date. Local structure has only been investigated in related
CaCu$_3$Ru$_4$O$_{12}$, and in a different
context.\cite{ebbin;jssc02} Here we present a joint x-ray and neutron
diffraction study where the data are analyzed both in real and reciprocal
space.

\begin{figure}[tb]
%\begin{center}$\,$
\begin{center}
\epsfxsize=3.2in
\epsfbox{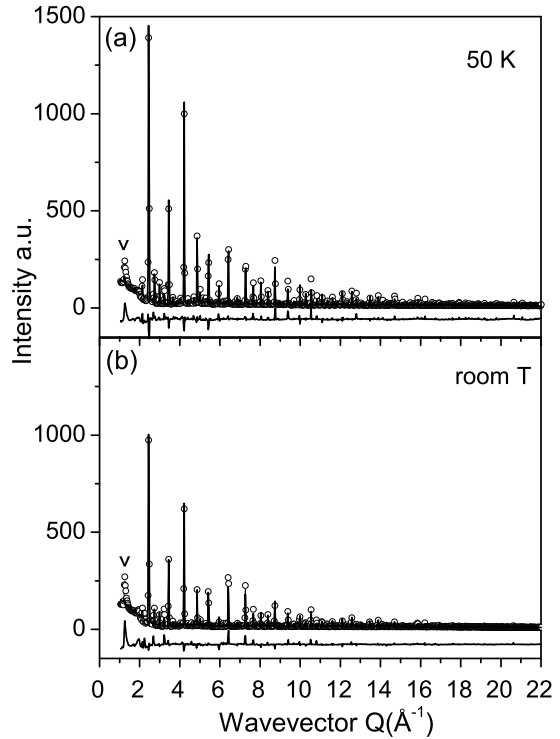}
\end{center}
\caption{Experimental (open circles) and Rietveld fitted X-ray diffraction
patterns (solid line) of CaCu$_3$Ti$_4$O$_{12}$ at (a) 50~K and (b) room
temperature. The residual difference is given in the bottom part of
each panel. The broad peak at very low Q-values (marked with V)
originates from the sample environment (aluminum shroud of the displex
cooling system).}
\protect\label{fig;rietveldfit}
\end{figure}

Typical Rietveld fits of the X-ray diffraction data are shown
in Fig.~\ref{fig;rietveldfit}.
%%%%%%%%%%%%%%%%%%%%%%%%%%%%%%%%%%%%%%%%%%%%%%%%%%%%%%%%%%%%%%%%%%
%  FIGURE
%%%%%%%%%%%%%%%%%%%%%%%%%%%%%%%%%%%%%%%%%%%%%%%%%%%%%%%%%%%%%%%%%%
%%%%%%%%%%%%%%%%%%%%%%%%%%%%%%%%%%%%%%%%%%%%%%%%%%%%%%%%%%%%%%%%%%
%  FIGURE
%%%%%%%%%%%%%%%%%%%%%%%%%%%%%%%%%%%%%%%%%%%%%%%%%%%%%%%%%%%%%%%%%%
The data are well explained within the $\im3$ space group at both low
and high temperature, confirming the absence of phase transitions
and distortions in the average structure, in quantitative agreement
with earlier results.\cite{subra;jssc00}

\begin{figure}[tb]
\begin{center}$\,$
\epsfxsize=3.2in
\epsfbox{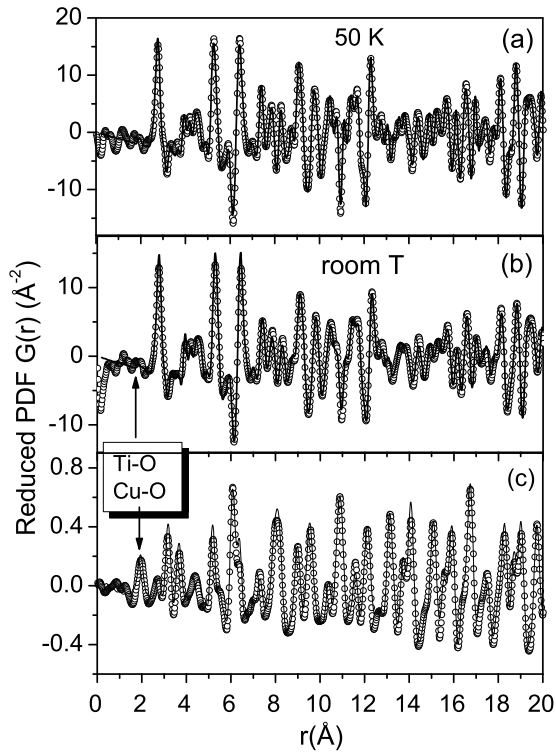}
\end{center}
\caption{Reduced atomic PDF of CCTO. Experimental data (open circles)
and fits (solid lines) shown for: (a) data obtained from neutron diffraction
at 50~K and (b) at room temperature; (c) data obtained from X-ray diffraction
at room temperature. Goodness of the PDF fit is around 15\%. The PDF profiles
corresponding to the two probes have different intensity distributions,
reflecting different scattering properties of the sample for neutron
diffraction and X-ray diffraction. Arrows mark the nearest neighbor PDF peak:
in CCTO it contains two contributions, coming from Ti-O and Cu-O distances
(see text).
}
\protect\label{fig;pdfprofile}
\end{figure}

Typical PDFs of CCTO both from neutron diffraction and X-ray
diffraction experiments are shown in Fig.~\ref{fig;pdfprofile}.
%%%%%%%%%%%%%%%%%%%%%%%%%%%%%%%%%%%%%%%%%%%%%%%%%%%%%%%%%%%%%%%%%%
%  FIGURE
%%%%%%%%%%%%%%%%%%%%%%%%%%%%%%%%%%%%%%%%%%%%%%%%%%%%%%%%%%%%%%%%%%
%%%%%%%%%%%%%%%%%%%%%%%%%%%%%%%%%%%%%%%%%%%%%%%%%%%%%%%%%%%%%%%%%%
%  FIGURE
%%%%%%%%%%%%%%%%%%%%%%%%%%%%%%%%%%%%%%%%%%%%%%%%%%%%%%%%%%%%%%%%%%
It should be noted that {\it neutron} PDFs of CCTO exhibit peculiar behavior:
contributions of Ti-O and Cu-O to the first PDF peak cancel out, so that the
first peak is `missing'. This is  due to the negative scattering length of Ti
for neutrons, and due to the almost identical Ti-O and Cu-O distances. 
This perfect cancellation of the Cu-O and Ti-O intensities (e.g.\ 
compare Fig.~\ref{fig;pdfprofile} (a), (b) and (c) at $r=1.9$~\AA) will only 
occur for selected peaks in the structure. Other peaks contain information 
about the relative positions of these ions and so a refinement over a 
wide-range of $r$-will provide accurate structural information on these 
species.

Structural modeling of the CCTO PDF profiles was performed observing
constraints of the $\im3$ symmetry. Structural parameters are summarized
in Table~\ref{tab1;structuralresults}. The lattice parameter shows the
expected temperature dependence. 
%%%%%%%%%%%%%%%%%%%%%%%%%%%%%%%%%%%%%%%%%%%%%%%%%%%%%%%%%%%%%%%%%%
%  FIGURE
%%%%%%%%%%%%%%%%%%%%%%%%%%%%%%%%%%%%%%%%%%%%%%%%%%%%%%%%%%%%%%%%%%
\begin{table}[tb]
\caption{Results of the neutron powder diffraction structural
study of CCTO, observing $\im3$ space group constraints:
lattice parameter and atomic coordinates. The atomic positions
are Ca(0,0,0), Cu(0,0.5,0.5), Ti(0.25,0.25,0.25), O({\it x},{\it y},0). 
Numbers in parentheses are expected standard deviations obtained from fitting.
}
\label{tab1;structuralresults}
\begin{tabular}{cccc}
T (K)&a (\AA)&{\it x}(O)&{\it y}(O) \\
\colrule
 50&7.3818(5)&0.3029(1)&0.1790(1) \\
100&7.3823(5)&0.3029(1)&0.1789(1) \\
150&7.3836(5)&0.3031(1)&0.1790(1) \\
200&7.3856(5)&0.3031(1)&0.1791(1) \\
290&7.3906(5)&0.3032(1)&0.1792(1) \\
\end{tabular}
\end{table}
%%%%%%%%%%%%%%%%%%%%%%%%%%%%%%%%%%%%%%%%%%%%%%%%%%%%%%%%%%%%%%%%%%
%  FIGURE
%%%%%%%%%%%%%%%%%%%%%%%%%%%%%%%%%%%%%%%%%%%%%%%%%%%%%%%%%%%%%%%%%%%
Corresponding isotropic displacement parameters, 
U$_{iso}\equiv\langle$u$^2_{iso}\rangle$ in standard notation, are
listed in Table~\ref{tab2;displacementparameters}.
%%%%%%%%%%%%%%%%%%%%%%%%%%%%%%%%%%%%%%%%%%%%%%%%%%%%%%%%%%%%%%%%%%
%  FIGURE
%%%%%%%%%%%%%%%%%%%%%%%%%%%%%%%%%%%%%%%%%%%%%%%%%%%%%%%%%%%%%%%%%%
\begin{table}[tb]
\caption{Isotropic displacement parameters, U$_{iso}$, in units
of \AA$^2$, as obtained from PDF structural refinements of neutron
powder diffraction data. Numbers in parentheses are standard
deviations estimated from fitting.
}
\label{tab2;displacementparameters}
\begin{tabular}{ccccc}
T (K)&Ca&Cu&Ti&O \\
\colrule
 50&0.0041(2)&0.0027(2)&0.0028(2)&0.0034(2) \\
100&0.0048(2)&0.0028(2)&0.0032(2)&0.0038(2) \\
150&0.0055(2)&0.0033(2)&0.0034(2)&0.0040(2) \\
200&0.0060(2)&0.0042(2)&0.0039(2)&0.0043(2) \\
290&0.0057(2)&0.0056(2)&0.0047(2)&0.0050(2) \\
\end{tabular}
\end{table}
%%%%%%%%%%%%%%%%%%%%%%%%%%%%%%%%%%%%%%%%%%%%%%%%%%%%%%%%%%%%%%%%%%
%  FIGURE
%%%%%%%%%%%%%%%%%%%%%%%%%%%%%%%%%%%%%%%%%%%%%%%%%%%%%%%%%%%%%%%%%%
Nearest neighbor distances are shown in Fig.~\ref{fig;bondlengths}:
%%%%%%%%%%%%%%%%%%%%%%%%%%%%%%%%%%%%%%%%%%%%%%%%%%%%%%%%%%%%%%%%%%
%  FIGURE
%%%%%%%%%%%%%%%%%%%%%%%%%%%%%%%%%%%%%%%%%%%%%%%%%%%%%%%%%%%%%%%%%%
\begin{figure}[tb]
\begin{center}$\,$
\epsfxsize=3.2in
\epsfbox{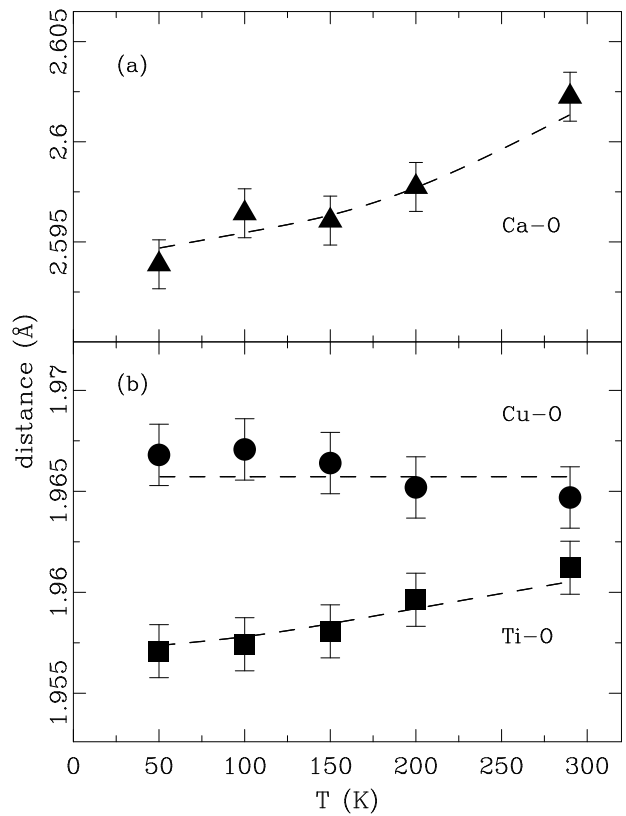}
\end{center}
\caption{Temperature dependence of the nearest neighbor distances
in CaCu$_3$Ti$_4$O$_{12}$: (a) Ca-O (solid triangles), (b) Cu-O
(solid circles) and Ti-O (solid squares). Dashed lines are guides
for the eye.
}
\protect\label{fig;bondlengths}
\end{figure}
%%%%%%%%%%%%%%%%%%%%%%%%%%%%%%%%%%%%%%%%%%%%%%%%%%%%%%%%%%%%%%%%%%
%  FIGURE
%%%%%%%%%%%%%%%%%%%%%%%%%%%%%%%%%%%%%%%%%%%%%%%%%%%%%%%%%%%%%%%%%%
while Ca-O (solid triangles) and Ti-O (solid squares) distances
decrease monotonically with decrease in temperature, Cu-O distance
(sold circles) appears to be temperature independent, reflecting the
rigidity of the square-planar coordination of Cu sublattice. Rietveld
obtained distances (not shown) are in agreement with PDF, within the
experimental uncertainties. The distances at room temperature are in
good agreement with those reported in Ref.~\onlinecite{bochu;jssc79}, but
differ somewhat from those reported in Ref.~\onlinecite{subra;jssc00} where
the temperature dependence of the Cu-O bond lengths appears to be over
estimated. 

The results of the polyhedral analysis revealed very small changes in
the volumes and shapes of TiO$_6$ octahedra on cooling from room
temperature down to 50~K. Changes in the magnitude of the octahedral
tilting, as expressed by the Ti-O-Ti angle, are smaller than 0.3$^{\circ}$
(less than 0.2\% change).

\section{DISCUSSION}

\subsection{Evidence for non-Debye local atomic displacements}

\begin{figure}[tb]
\begin{center}$\,$
\epsfxsize=3.2in
\epsfbox{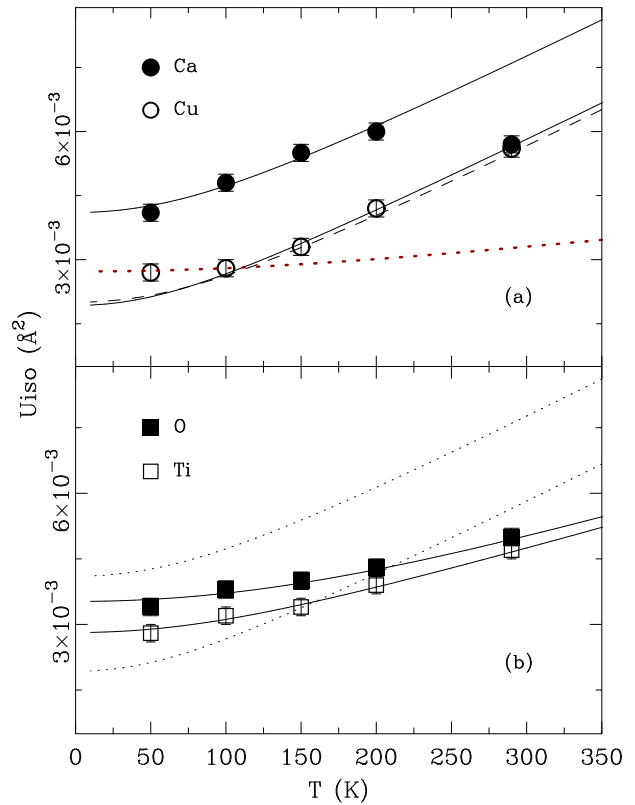}
\end{center}
\caption{(a) Evolution with temperature of the isotropic displacement
parameters of Ca and Cu. Solid lines are from the Debye model for Ca and Cu, 
with parameters as specified in {\protect Table~\ref{tab3;debye}}. Dashed line
is a Debye curve of Ca corresponding to the same Debye temperature, but
with offset parameter set to zero (see text). Thick dotted line is a Debye
model for Cu using higher Debye temperature of 785~K. (b) Isotropic
displacement parameters of Ti and O as a function of temperature, with Debye
model curves superimposed (solid lines). Debye model curves for Ca and Cu
(dotted lines) are also shown for comparison.
}
\protect\label{fig;results}
\end{figure}

Inspection of the PDF displacement parameters, that provide clues on local
disorder, revealed an unusual temperature dependence of the isotropic
displacement parameter of calcium. At low temperature it is 50\% larger 
than the corresponding value for copper.  However, at room temperature the two
appear to be equal (Fig.~\ref{fig;results}). We should note that this observed 
behavior relies on a single high-temperature data-point.  We are in the 
process of extending our data collection in this region to confirm this 
result and extract in greater detail the temperature dependence of this 
parameter.

The data have been fit with a Debye model to understand whether or not the
low-temperature displacement factors are
anomalous.\cite{kitte;bk63,beni;prb76,jeong;prb03} Two-parameter fitting
of the Debye curve to the displacement parameter temperature dependencies
has been carried out, using
\begin{equation}
\sigma^2_D= \frac{3\hbar^2}{mk_B\Theta_D}\left[\frac{1}{4}+\left(\frac{T}{\Theta_D}
\right)^2\int_{0}^{\frac{\large\Theta_D}{T}}\frac{x}{e^x-1}dx\right]+\sigma^2_{off},
\label{eq;debyeequation}
\end{equation}
where the parameters are the Debye temperature, $\Theta_D$, and the
static offset, $\sigma^2_{off}$. Results of the fitting are summarized
in Table~\ref{tab3;debye}
%%%%%%%%%%%%%%%%%%%%%%%%%%%%%%%%%%%%%%%%%%%%%%%%%%%%%%%%%%%%%%%%%%
%  FIGURE
%%%%%%%%%%%%%%%%%%%%%%%%%%%%%%%%%%%%%%%%%%%%%%%%%%%%%%%%%%%%%%%%%%
\begin{table}
\caption{Estimate of the Debye temperature, $\theta_D$, and static offset
parameter, $\sigma^2_{off}$, for various constituent atoms of CCTO.
Parameters are obtained by fitting Debye model to the temperature
dependencies of the PDF displacement amplitudes.
}
\label{tab3;debye}
\begin{tabular}{ccc}
Atom&$\Theta_D$ (K)&$\sigma^2_{off}$ (\AA$^2$) \\
\colrule
 Ca&450(20)&0.0021(1) \\
 Cu&360(20)&0.0004(9) \\
 Ti&545(20)&0.0014(2) \\
 O&923(20)&0.0011(1)    \\
\end{tabular}
\end{table}
%%%%%%%%%%%%%%%%%%%%%%%%%%%%%%%%%%%%%%%%%%%%%%%%%%%%%%%%%%%%%%%%%%
%  FIGURE
%%%%%%%%%%%%%%%%%%%%%%%%%%%%%%%%%%%%%%%%%%%%%%%%%%%%%%%%%%%%%%%%%%
and shown in Fig.~\ref{fig;results}.
%%%%%%%%%%%%%%%%%%%%%%%%%%%%%%%%%%%%%%%%%%%%%%%%%%%%%%%%%%%%%%%%%%
%  FIGURE
%%%%%%%%%%%%%%%%%%%%%%%%%%%%%%%%%%%%%%%%%%%%%%%%%%%%%%%%%%%%%%%%%%
%%%%%%%%%%%%%%%%%%%%%%%%%%%%%%%%%%%%%%%%%%%%%%%%%%%%%%%%%%%%%%%%%%
%  FIGURE
%%%%%%%%%%%%%%%%%%%%%%%%%%%%%%%%%%%%%%%%%%%%%%%%%%%%%%%%%%%%%%%%%%

Oxygen and Ti parameters exhibit the expected behavior with relatively
high Debye temperatures and small offsets. The fits are shown in
Fig.~\ref{fig;results}(b). The large Debye temperature of oxygen is
similar to that obtained in cubic manganites.\cite{proff;prb99} The Debye 
parameters for Ti are also similar to perovskite titanates such as 
BaTiO$_3$.\cite{kwei;jpc93} There is, therefore, no evidence for local
disorder on the titanium or oxygen sublattices. This suggests that
nano-scale tilt disorder, as envisioned by Ramirez~\etal \cite{ramir;cm02}
to explain colossal dielectric behavior, is not present.

The Ca and Cu parameters behave more interestingly. The atomic displacement
parameter associated with the Ca sites remains unusually large at low
temperature (50\% higher than that of Cu) with a significant
offset parameter. This may suggest local static displacive disorder on this
site, with a magnitude of approximately 0.05~\AA . This will be discussed
later in terms of bond-valence sum predictions for the Ca site. We also note
that all the points except the high temperature one are well explained by
the fit. The dashed line in Fig.~\ref{fig;results}(a) is the same Debye curve
of Ca, but with the offset parameter {\it set to zero}. It is interesting
to note that the high temperature U$_{iso}$(Ca) point lies on the dashed
curve where the static component is switched off. This observation may be
suggestive of a crossover from a disordered state at low temperature to a
more ordered state at high temperature, with crossover temperature in the
200-300~K range. Further measurements to verify this result are planned.

Also interesting is the low Debye temperature obtained for copper from the
fitting. Copper is expected to be stiffly bonded to its four close oxygen
neighbors and a much higher Debye temperature than the observed 360~K might 
be expected. It is also notable that the Debye fit (Fig.~\ref{fig;results}(a) 
lower solid line) fails at low temperature for the copper site with
the low-temperature data-point displaced upwards from the Debye curve.
A possible explanation is that the actual Debye temperature of the copper
is actually higher than indicated by the fit. This would give a flatter Debye
curve and explain the low-T behavior, as indicated by a thick dotted line
in Fig.~\ref{fig;results}(a) corresponding to a Debye temperature of 785~K.
In this picture, above 100~K an additional non-Debye but T-dependent excess
disorder appears on this site.

\subsection{Bond valence sum prediction of the crystal structure of CCTO}
Here we discuss bond valence sum calculations that make predictions about
the stability as a function of temperature of the observed perovskite 
structure. These may indicate a tendency to local structural distortions 
that are prevented from propagating by the bracing described above.

Normally in an ABO$_3$ perovskite ($\it i.e.$ CaTiO$_3$) the
octahedral tilting distortion increases as the temperature
decreases.  This occurs because the A-O bonds contract
more rapidly in response to the decreasing temperature than
the B-O bonds. However, in CCTO the situation is somewhat
different.  Firstly, the Cu-O bond length is very sensitive to
changes in the magnitude of the tilting distortion, whereas the Ca-O
bond length contracts much more slowly with increasing tilt. Secondly,
the Cu-O bonds are considerably more covalent than the typical A-O bond,
and exhibit relatively little thermal expansion over the
temperature range examined in this study, as shown in
Fig.~\ref{fig;bondlengths}. These considerations help to
explain why both the tilt magnitude and the Cu-O distance are
essentially independent of temperature.

\begin{figure}[tb]
\begin{center}$\,$
\epsfxsize=3.2in
\epsfbox{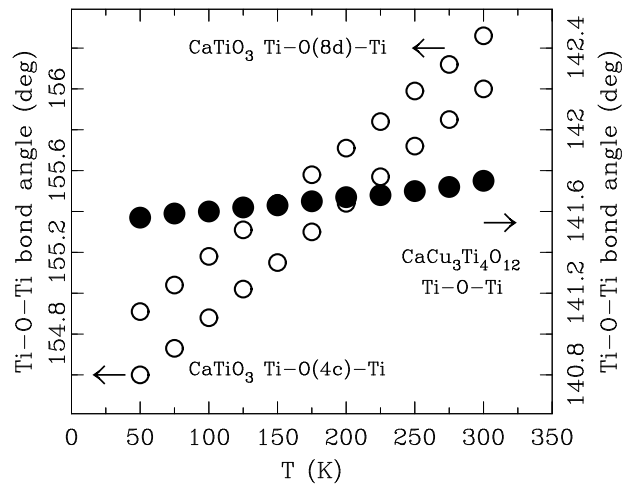}
\end{center}
\caption{Expected Ti-O-Ti bond angles as a function of temperature for
CaCu$_3$Ti$_4$O$_{12}$ and CaTiO$_3$ from bond valence analysis.}
\protect\label{fig;tiotibondangle}
\end{figure}

\begin{figure}[tb]
\begin{center}$\,$
\epsfxsize=3.2in
\epsfbox{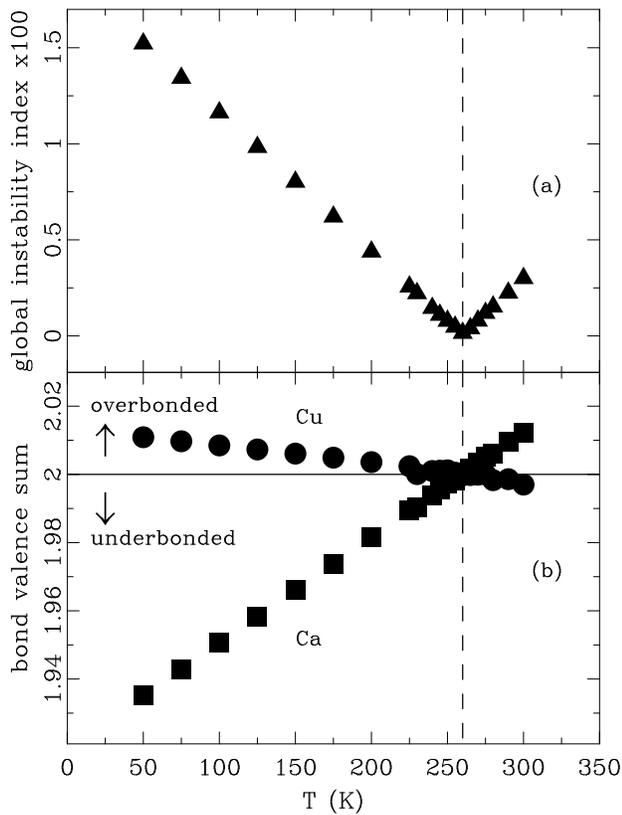}
\end{center}
\caption{(a) Global instability index (GII) as a function of temperature,
calculated for CaCu$_3$Ti$_4$O$_{12}$. (b) Predicted bond valence sums
for Ca and Cu. Solid line denotes optimal valence for the two atomic types,
while dashed line marks minimum of the GII where the bonding requirements
of Ca and Cu are optimally matched.}
\protect\label{fig;giibvs}
\end{figure}

As described above, the Cu-O bonding locks-in the tilting at what
is essentially a constant value. However, the optimal Ca-O bond length
should still contract fairly substantially with decreasing temperature,
as it would in a typical perovskite such as CaTiO$_3$. Consequently,
there will be a crossover temperature where the Ca-O and Cu-O bonding
are perfectly matched. Above this temperature the calcium will be
overbonded. That is to say that the effective size of Ca will be too
large for its environment and the Ca-O bonds will be in compression.
Below the optimal temperature the calcium will be underbonded, so that
the effective size of Ca is too small for its environment and the Ca-O
bonds will be in tension. In order to estimate the optimal temperature
for CCTO we have utilized the structure prediction capabilities of the
SPuDS software package\cite{lufaso;actab01} to evaluate the competition
between Cu-O, Ca-O and Ti-O bonding as a function of temperature.
SPuDS is based on the bond valence concept\cite{brown;csr78,brown;acb85}
and has been shown to be quite accurate in predicting the magnitude of
octahedral tilting distortions in perovskites.\cite{lufaso;actab01}
The thermal expansion of all bonds was assumed to be inversely related
to the valence of the bond as suggested by Brown.\cite{brown;acb97}
Figure~\ref{fig;tiotibondangle}
%%%%%%%%%%%%%%%%%%%%%%%%%%%%%%%%%%%%%%%%%%%%%%%%%%%%%%%%%%%%%%%%%%
%  FIGURE
%%%%%%%%%%%%%%%%%%%%%%%%%%%%%%%%%%%%%%%%%%%%%%%%%%%%%%%%%%%%%%%%%%
%%%%%%%%%%%%%%%%%%%%%%%%%%%%%%%%%%%%%%%%%%%%%%%%%%%%%%%%%%%%%%%%%%
%  FIGURE
%%%%%%%%%%%%%%%%%%%%%%%%%%%%%%%%%%%%%%%%%%%%%%%%%%%%%%%%%%%%%%%%%%
shows the predicted Ti-O-Ti angles as a function of temperature for
CaCu$_3$Ti$_4$O$_{12}$ and CaTiO$_3$. In agreement with the experimental
results reported here the temperature dependence of the tilting
in CCTO is predicted to be much smaller than that of CaTiO$_3$.
The predicted calcium and copper bond valence sums, as well as the
global instability index (GII),\cite{salin;jssc92} are shown in
Fig.~\ref{fig;giibvs}
%%%%%%%%%%%%%%%%%%%%%%%%%%%%%%%%%%%%%%%%%%%%%%%%%%%%%%%%%%%%%%%%%%
%  FIGURE
%%%%%%%%%%%%%%%%%%%%%%%%%%%%%%%%%%%%%%%%%%%%%%%%%%%%%%%%%%%%%%%%%%
%%%%%%%%%%%%%%%%%%%%%%%%%%%%%%%%%%%%%%%%%%%%%%%%%%%%%%%%%%%%%%%%%%
%  FIGURE
%%%%%%%%%%%%%%%%%%%%%%%%%%%%%%%%%%%%%%%%%%%%%%%%%%%%%%%%%%%%%%%%%%
as a function of temperature. The global instability index reaches a
minimum near 260~K, where the bonding requirements of calcium and copper
are optimally matched. Below this temperature the Ca atoms become
underbonded and the copper atoms overbonded.  Even though this method
of estimating the crossover temperature is fairly crude it is interesting to
note that what seems to be a crossover in the Ca isotropic displacement
parameter takes place near this temperature.

\subsection{Local structure and braced lattice model}

Despite the large tilting of TiO$_6$ octahedra in CCTO, no structural
phase transition to a lower symmetry phase, as observed in related
CaTiO$_{3}$, SrTiO$_{3}$ and
BaTiO$_{3}$,\cite{redfe;jpcm96,tsuda;aca95,kwei;jpc93} is evident.
These transitions typically occur because underbonded ions
(i.e., ions that are too small for their coordination environment)
desire to lower their site symmetry.  This should be observed in the
current case, because Ca is in this underbonded state at low temperature. 
However, no symmetry lowering phase transitions are observed. This has 
recently been argued to be due to the bracing of the perovskite structure 
by the Cu-O square complex preserving the cubic structure of CCTO at all 
temperatures.\cite{ramir;cm02} Distorting the Cu-O bonds is energy expensive, 
and this prevents changes in TiO$_6$ octahedral tilting, as such changes 
would affect the copper coordination by changing its oxygen-square 
environment. Rigidity of the Cu-O complex is evident from the temperature 
independence of the Cu-O nearest neighbor distance 
(Fig.~\ref{fig;bondlengths} (b)). Our displacement parameter results described 
above suggest that it is plausible that off-center displacements of Ca may be 
present in the structure that are not long-range ordered and do not result in 
a phase transition.

Ramirez and collaborators investigated, within the mean-field
approximation, a possibility of having isolated polarizable
defects in otherwise ideal CCTO structure.\cite{ramir;cm02}
It is speculated that the defects are either Cu vacancies or
interchange of Ca and Cu sites in such a manner that the rigid
square Cu-O network, bracing the structure, is {\it locally}
disrupted. This local absence of bracing would, according to this
scenario, allow for local, short-range ordered, distortions to take place.
The lack of average structural evidence is argued to be due to the
low concentration of defect sites, as well as due to the relaxation
between alternative equivalent configurations that would preserve
the average cubic structure. It is then expected that a diffuse
scattering signal in neutron scattering experiments would be observable
in the large-dielectric-constant regime.  Polyhedral analysis,
as mentioned above, does not reveal changes in the octahedral size,
shape, and rotation in the local structure of CCTO.  This argues against
locally correlated distortions of the octahedral network.  On the other hand,
we do see evidence for possible disorder on the Ca and Cu sublattices that
may be locally correlated.  Observing possible local correlations in Ca
displacements is beyond the accuracy of our measurement.

Such polarizable defects, embedded into a background with dielectric
constant $\epsilon_0$ and relaxing at a rate determined by temperature
and an energy barrier separating alternative configurations, are proposed
to have small concentration, of the order of 10$^{-3}$. The detection of 
defects at this level is beyond our measurements. However, the
idea that locally interrupted bracing due to these defects would allow for 
local distortions involving TiO$_{6}$ units to take place is inconsistent 
with the PDF results, and also the results of recent polarized Raman study 
on single-crystal CCTO and within broad temperature range.\cite{kolev;prb02} 
Therefore, local octahedral tilting distortions can be clearly ruled out as a 
candidate for the observed large dielectric constant.

\subsection{Implications for dielectric properties}

We address possible implications of the disorder on Ca and Cu
sublattices for the dielectric properties of CCTO. Although there 
is now mounting evidence that the unusual dielectric properties of 
CCTO may be extrinsic in nature,\cite{lunke;prb02,subra;sss02} the 
possibility for an intrinsic mechanism is not completely ruled 
out.\cite{ramir;cm02} Even with an extrinsic mechanism the question 
remains what is special about CCTO and why is such dielectric behavior 
not much more widely observed. One of the remaining plausible avenues 
for the intrinsic mechanism involves a system that is at the edge of 
the Clausius-Mossotti catastrophe.\cite{lunke;prb02} Origin of the 
giantic dielectric constant in the low-frequency mode and also unusually 
large dielectric constant in the high-frequency regime (of the order of
100), as well as the exact nature of the switching mechanism
between these two are not clearly understood.

The defect-model in Ref.~\onlinecite{ramir;cm02} describes a system
involving a small concentration of polarizable defects, with a
temperature independent polarization, embedded in a background
dielectric constant, $\epsilon_{0}$. The enormous real part of the 
dielectric constant of such a system is then readily achieved providing 
(i) $\epsilon_0 \gg 1$, (ii) the polarizability of the defects is of the 
order of their volume, and (iii) the concentration of the defects is 
reasonably small, $\ll 1$. This model appears to capture rather nicely 
the unusually high value of the measured dielectric constant plateau of 
the order of 5000.\cite{ramir;cm02} However, sensitivity of the dielectric
constant plateau value, within this model, to the exact value of
$\epsilon_0$ used is remarkable. Changing $\epsilon_0$ from 100 to
about 70, and keeping all other parameters in the model the same,
changes the plateau value from about 5128 to only about 223. This
demonstrates the importance of the background dielectric constant
within this model.

Study of the optical properties of CCTO\cite{homes;sci01} reveals
an unusual temperature dependence of the real part of the
dielectric function in the far-infrared (FIR) limit ($\omega \le$~20 cm$^{-1}$)
, $\epsilon_{FIR}$. This is found to change from
$\sim 70$ at high temperature, to $\sim 120$ at base temperature,
an increase of about 70\%. The order of ${\large
\epsilon}_{\small FIR}$ was concluded to be in good agreement with
the high frequency limit (and low temperature) value of the
measured dielectric constant of CCTO, presumably $\epsilon_0$
featured in the defect-model. The increase in the FIR value of the
dielectric constant has an onset temperature of the order of
200~K, and its temperature dependence has been assigned to the
violation of the $f$-sum rule and the increase of the oscillator
strengths of the IR modes. This onset temperature seems to
correlate well with the onset temperature of the local static
disorder on Ca sublattice observed from the PDF analysis.

While the issues addressed here are far from being resolved, we
speculate that Ca sublattice may play a key-role in explaining
some of the observed properties of CCTO. Distribution of the Ca
atoms between displaced lattice sites in some sort of multi-well
potential at low temperatures may provide a natural explanation
for the observed enlarged thermal amplitudes. Furthermore, if this
happens in a {\it locally} correlated fashion, within some sort of
nanodomain structure, it could possibly be relevant for the
anomalous increase of far-infrared $\epsilon_{FIR}$ observed in
studies of the optical properties.\cite{homes;sci01} In this view,
upon increasing temperature, the local order on Ca sites is
destroyed, and this leads to the decrease of the intrinsic (high
frequency) dielectric constant. Small volume fraction polarizable defects, 
be they intrinsic or extrinsic, would then give the required amplification 
of the low-frequency dielectric constant through the mechanism described 
in Ref.~\onlinecite{ramir;cm02}.

\section{SUMMARY}

Local structural properties of the CCTO system are investigated as
a function of temperature within a range from 50~K up to room
temperature. Data were analyzed using Rietveld refinement and the
atomic pair distribution function (PDF) analysis of the powder
diffraction data has been employed to extract the local structural
information. In agreement with earlier studies, no signature of a
phase change has been observed. The temperature dependence of the
nearest neighbor Cu-O distance is very flat, supporting the idea
that this is bracing the structure. No evidence for
perovskite-type local distortions is found. The isotropic Ca
displacement parameter has a significant static offset compared to
the expected Debye behavior. It also appears to behave anomalously
at high-temperature, though this needs to be verified by more
measurements.  The Debye temperature of copper is surprisingly
low, given its covalently bonded nature. We note that the
T-dependence of the Cu displacement parameters could be reconciled
with a higher Debye temperature coupled with non-Debye but temperature
dependent displacements, where these extra displacements begin
appearing above $\sim 100$~K.

Simple modeling based on bond valence concepts supports the local
structural observations. The temperature at which both the Ca and
Cu environments are optimal has been estimated to be in the
vicinity of 260~K. At lower temperatures the Ca becomes underbonded. 
This may result in an off-center displacement of Ca ions. If these 
displacements cannot become long-range ordered because of the braced 
structure, this might explain the absence of a structural phase 
transition and the static offset of the Ca thermal parameters.

Local domains of coherently polarized displaced Ca ions could
possibly exist at low temperature and may be related to the
enhanced values of far-IR dielectric constant, a parameter playing
an important role in intrinsic-mechanism models for the large
dielectric response at the edge of dielectric catastrophe, such as
that proposed by Ramirez~\etal\cite{ramir;cm02}. This model seems
capable of explaining the large dielectric constants; however, it
is very sensitive to the exact value of the background far-IR
dielectric response of the material that provides an amplification
effect. From this study we cannot say whether such an intrinsic
mechanism, or an extrinsic mechanism, is responsible for the
colossal dielectric behavior.

\section{ACKNOWLEDGEMENTS}
We are thankful to Simine Short for providing valuable help with
the neutron diffraction data collection. We would like to thank
Mike Lufaso for his development of the program SPuDS. This work
was supported financially by DOE-BES through grant FG02-97ER45651.
The experimental data were collected on X7A beamline at the NSLS
at Brookhaven National Laboratory, and on SEPD at the IPNS at
Argonne National Laboratory. NSLS is funded by the US Department of 
Energy under contract DE-AC02-98CH10886. IPNS is funded by the US 
Department of Energy under contract W-31-109-ENG-38. 

%---------------------------------------------------------------------------
% References
% NOTE: Add .bbl file here before submitting the file !!!!!!!!!!!!!!!!!!!!!!
%---------------------------------------------------------------------------

%\bibliography{\bibpath{emilCCTO2002}}
%\bibliographystyle{\bibpath{aip_simon}}

\end{document}